%% file: main.tex
\documentclass[]{pasj02} 
\usepackage[switch,mathlines]{lineno} 
\usepackage{natbib} 
\usepackage{url} 

\jyear{2025}
\Accepted{2025/08/08}

\usepackage{xspace}
\usepackage{color}

\newcommand\I{3I/ATLAS\xspace}
\newcommand\gr{$g-r=0.603\pm0.031$\xspace}
\newcommand\ri{$r-i=0.210\pm0.031$\xspace}
\newcommand\iz{$i-z=0.117\pm0.046$\xspace}
\newcommand\rz{$r-z=0.327\pm0.035$\xspace}
\newcommand\grval{$0.603\pm0.031$\xspace}
\newcommand\rival{$0.210\pm0.031$\xspace}
\newcommand\izval{$0.117\pm0.046$\xspace}
\newcommand\rzval{$0.327\pm0.035$\xspace}

\begin{document} 

\title{Simultaneous visible spectrophotometry of interstellar object 3I/ATLAS with Seimei/TriCCS}

\author{
Jin \textsc{Beniyama}\altaffilmark{1,2}\altemailmark\orcid{0000-0003-4863-5577} \email{jbeniyama@oca.eu} 
}
\altaffiltext{1}{Université Côte d’Azur, Observatoire de la Côte d’Azur, CNRS, Laboratoire Lagrange, Bd de l’Observatoire, CS 34229, 06304 Nice Cedex 4, France}
\altaffiltext{2}{The Department of Earth and Planetary Science, The University of Tokyo, 7-3-1 Hongo, Bunkyo, Tokyo 113-0033, Japan}


\maketitle

\begin{abstract}
3I/ATLAS, also known as C/2025 N$_1$ (ATLAS), is the third interstellar object (ISO) discovered in July 2025.
ISOs are particularly interesting because characterizing their physical properties helps us understand and test our knowledge of Solar System formation.
Several quick response observations of 3I/ATLAS were performed during the first few days after the discovery, and various results, such as reflectance spectra, have been reported.
We performed simultaneous visible spectrophotometry of 3I/ATLAS from data taken using the TriColor CMOS Camera and Spectrograph (TriCCS) on the Seimei 3.8~m telescope.
The Seimei/TriCCS observations of 3I/ALTAS were obtained in the $g$, $r$, $i$, and $z$ bands in the Pan-STARRS system on UTC July 15, 2025.
Our lightcurves show no significant variations during the 2.3~h observation, which is in good agreement with previous studies.
Visible color indices of 3I/ATLAS, \gr, \ri, \iz, and \rz suggest it has a red surface similar to, or slightly redder than, that of D-type asteroids.
Continuous observations of 3I/ATLAS before and after its perihelion passage in October 2025 are desired to investigate its physical properties.
\end{abstract}

\clearpage

\section{Introduction}
Small bodies formed outside of the Solar System but are passing through it are called interstellar objects (ISO) or interstellar interlopers \citep[e.g.,][for review]{Jewitt2023}.
    ISOs are particularly of interest 
    because their physical characteristics can reveal valuable information about the early history of the Solar System.
    A comparative study of ISOs and Small Solar System Bodies (SSSBs) might allow us to investigate the planetesimal formation in different systems.
The first ISO 1I/`Oumuamua was discovered by the Pan-STARRS survey \citep{Chambers2016} on October 19, 2017 \citep{Meech2017}, while the second ISO 2I/Borisov was discovered by G. Borisov on August 30, 2019.
    The two ISOs were quickly and extensively characterized following their discoveries,
    as such observations would be nearly impossible once the ISOs leave the Solar System.
    The surface colors of ISOs at visible to near-infrared wavelengths give insight into their composition and potential origins by comparing them with those of SSSBs
    \citep[e.g.,][]{Meech2017, Ye2017_1I, Jewitt2017, Bannister2017, Bolin2018_1I, Fitzsimmons2018, Jewitt2019_2I, Guzik2020_2I, Bolin2020_2I}.
Since 2019, no ISOs had been discovered for about six years, and our knowledge of ISOs remained limited due to the small sample size.

On July 1, 2025, the third ISO 3I/ATLAS, also known as C/2025 N$_1$ (ATLAS), was discovered by the Asteroid Terrestrial-impact Last Alert System \citep[ATLAS,][]{Tonry2018} in Chile \citep{Denneau2025}.
Soon after the discovery, weak cometary activity was reported by several observers across the world.
Initial characterizations were performed worldwide using instruments such as the Canada-France-Hawaii Telescope  \citep{Seligman2025_3I}, the Very Large Telescope \citep{Seligman2025_3I, Opitom2025_3I, Alvarez-Candal2025_3I}, 
Kottamia Astronomical Observatory 1.88-m telescope, Palomar 200-inch telescope, Apache Point Observatory Astrophysical Research Consortium 3.5-m telescope \citep{Bolin2025_3I, Belyakov2025_3I}, the 10.4 m Gran Telescopio Canarias \citep{Marcos2025_3I}, the NASA Infrared Telescope Facility \citep[IRTF,][]{Kareta2025_3I, Yang2025_3I}, 
    the Gemini South telescope \citep{Yang2025_3I},
and several smaller telescopes.
Also, the 37 detections from the Vera C. Rubin Observatory were reported \citep{Chandler2025_3I}.
The results of existing observations are diverse, with different spectra reported, making interpretation challenging.
    Regarding its lightcurves, \citet{Marcos2025_3I} reported 
    a periodicity of $16.79\pm0.23$~hours based on lightcurves obtained over approximately three days, 
    while \citet{Seligman2025_3I} concluded that no periodicity is apparent from observations over a period of approximately four days.
Independent and reliable measurements are required before the perihelion passage in October 2025, when the comet could become more active.

In this letter, we present the results of simultaneous visible spectrophotometry of \I.
The observations were conducted on July 15, 2025, when \I was at a heliocentric distance of 4.03~au.
We describe the observations and data reduction in section \ref{sec:obs}, and present the results and discussion in section \ref{sec:res}.
We present the conclusion in section \ref{sec:conc}.

\section{Observations}\label{sec:obs}
Visible spectrophotometric data of \I were acquired on July 15, 2025.
A summary of the observational conditions is provided in table \ref{tab:obs}.
The predicted $V$ band magnitudes in table \ref{tab:obs} were obtained from NASA JPL Horizons \footnote{\url{https://ssd.jpl.nasa.gov/horizons}} using the Python package astroquery \citep{Ginsburg2019}.
Observations of \I were carried out with the TriColor CMOS Camera and Spectrograph (TriCCS) mounted on the 3.8~m Seimei Telescope \citep{Kurita2020} 
under a single-epoch DDT program (PI: Jin Beniyama).
The telescope is located at the Kyoto University Okayama Observatory (133.5967$^\circ$ E, 34.5769$^\circ$ N, and 355~m in altitude).
Three-band imaging was simultaneously conducted using combinations of the Pan-STARRS ($g$, $r$, $i$) and ($g$, $r$, $z$) filters.
The instrument covers a field of view of $12.6\arcmin \times 7.5\arcmin$ with a pixel scale of 0.350~arcsec~pixel$^{-1}$.

At the time, \I had a heliocentric distance of about 4.03~au and a geocentric distance of about 3.11~au.
The phase angle of \I was approximately 7.1~deg on July 15, 2025.
The apparent sky motion of \I was about 1.5~arcsec~min$^{-1}$.
The percent of the Moon illuminated was about 76\%, and the lunar elongation was about 87-88~deg.
The sky was clear, and the seeing measured by using in-field stars was 2.2--2.6~arcsec in the $r$ band.

We obtained images in the non-sidereal tracking mode during the observations of \I.
Also, we obtained sidereal tracking images to measure the seeing.
Exposure times were set to 60~s for all observations.
We performed standard image reduction, including bias subtraction, dark subtraction, and flat-fielding.
The astrometry of in-field sources from the Gaia Data Release 2 was performed using the astrometry.net software \citep{Lang2010}.

\begin{figure*}
 \begin{center}
 \includegraphics[width=1\hsize]{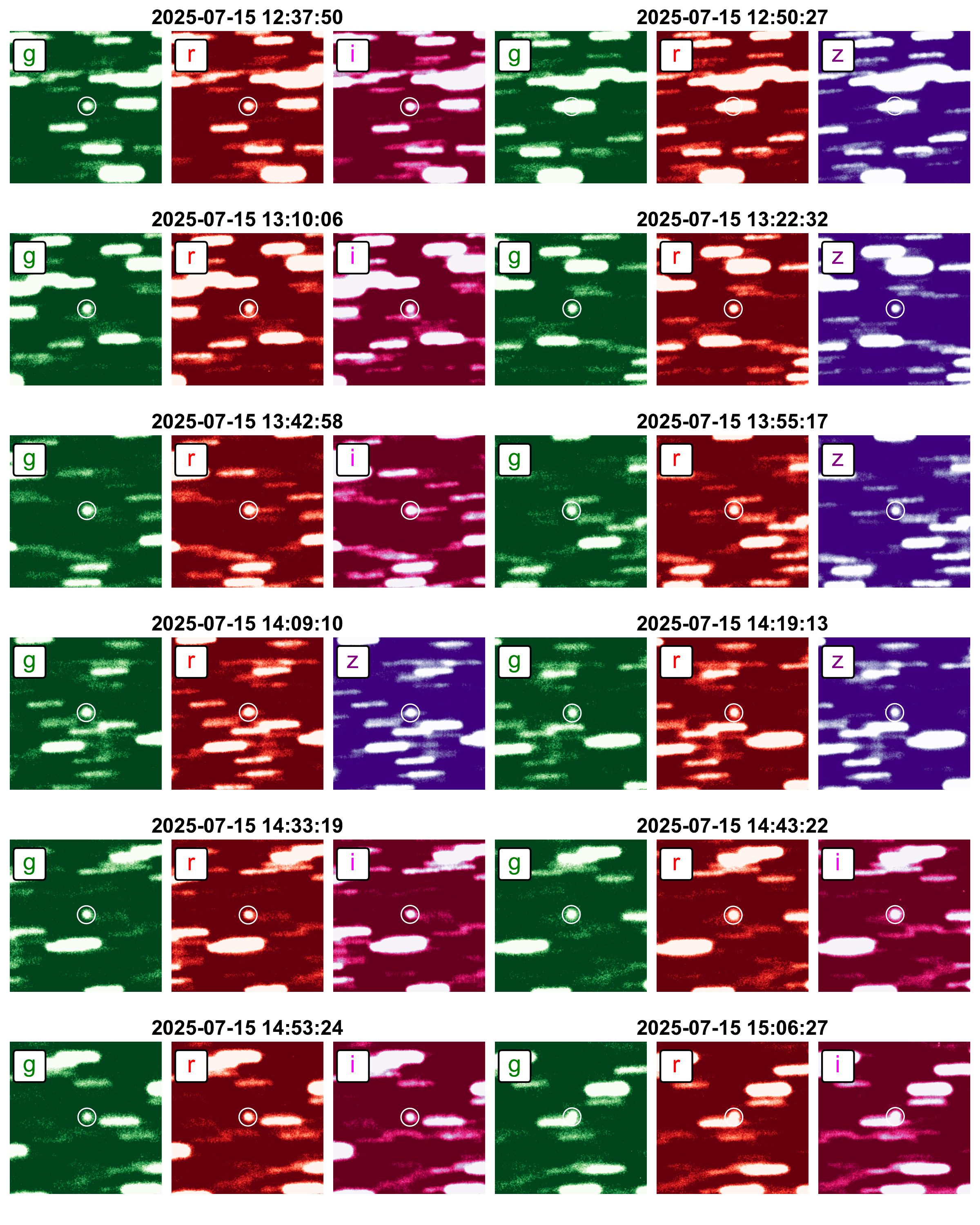}
 \end{center}
\caption{ Non-sidereally stacked frames
in $g$, $r$, $i$, and $z$ bands with a total integration time of 600~s on July 15, 2025 are shown.
Circles indicate \I.
Field of view covers $60~\arcsec\times60~\arcsec$.
North is to the top, and east is to the left.
 {Alt text: Non-sidereally stacked $g$, $r$, $i$, and $z$ band images of \I from July 15, 2025 (600 s total); 
$60~\arcsec\times60~\arcsec$ field of view; north is up, east is left; circles mark \I.}
}\label{fig:cutout}
\end{figure*}

To improve the signal-to-noise ratio (S/N) of \I while avoiding the elongations of their images, we performed mean stacking prior to photometry, as shown in figure \ref{fig:cutout}.
Ten successive 60~s exposures were stacked to obtain images with an effective exposure times of 600~s.
The two stacked frames, which overlap with the bright 
stars, were excluded from the analysis (see table \ref{tab:obs}).
A typical readout time of the CMOS sensors on TriCCS is 0.4~milliseconds,
which is negligibly small compared to the exposure time of 60~s.
We also stacked frames using the World Coordinate System (WCS) of images corrected with the surrounding sources to suppress the elongations of the images of reference stars.

A custom pipeline to perform photometry using the same instrument was developed and used in the previous papers \citep{Beniyama2023a, Beniyama2023b, Beniyama2023c, Beniyama2024, Beniyama2025b}.
The field of \I on that night was crowded with a galactic coordinate of $(l, b) = (7.7^{\circ}, 7.7^{\circ})$.
Thus, we developed and used a modified pipeline to analyze images of \I.
Circular aperture photometry was performed on both \I and the reference stars using the SExtractor-based Python package sep. 
The local background estimated as a median value of the $85\times85$~pix cutout image centered at the \I or a reference star was subtracted before photometry.
The aperture radii were set to 12~pix for reference stars, and were examined for \I (see section \ref{sec:res}).
Photometry of \I was conducted on non-sidereally stacked frames, while reference stars were measured using sidereal stacked frames.

All images were calibrated using Pan-STARRS catalog Data Release 2 \citep{Chambers2016}.
Since the field was crowded, we manually selected reference stars that are far from any other sources.
Extended sources, possible quasars as well as variable stars were removed using objinfoflag and objfilterflag in the Pan-STARRS catalog.
In total, 15 reference stars that met the following criteria were used in the analysis:
$0.0 \leq (g-r)_{\mathrm{PS}} \leq 1.1$ and $0.0 \leq (r-i)_{\mathrm{PS}} \leq 0.8$, where $(g-r)_{\mathrm{PS}}$ and $(r-i)_{\mathrm{PS}}$ are colors in the Pan-STARRS system.
The typical uncertainty in the magnitude zero point is approximately 0.05~mag

\begin{longtable}{rcccl}
\caption{Summary of the observations}\label{tab:obs}
\hline\noalign{\vskip3pt}
        Obs. Date & Filter & V     & Air Mass   & Note \\ [2pt]
        (UTC)     &        & (mag) &            & \\ [2pt]
\hline\noalign{\vskip3pt}
\endfirsthead
\hline\noalign{\vskip3pt}
        Obs. Date & Filter & V     & Air Mass   & Note \\ [2pt]
        (UTC)     &        & (mag) &            & \\ [2pt]
\hline\noalign{\vskip3pt}
\endhead
\hline\noalign{\vskip3pt}
\endfoot
\hline\noalign{\vskip3pt}
\multicolumn{2}{@{}l@{}}{\hbox to0pt{\parbox{160mm}{\footnotesize
\hangindent6pt\noindent
\hbox to6pt{\footnotemark[$*$]\hss}\unskip%
    Observation time in UT in midtime of exposure (Obs. Date), 
    and filters (Filter) are listed.\\
    Predicted V band apparent magnitudes (V)
    at the observation starting time are referred to \\
    NASA Jet Propulsion Laboratory (JPL) Horizons
     as of July 22, 2025.\\
    Elevations of 3I/ATLAS to calculate air mass range (Air Mass) are \\
    also referred to NASA JPL Horizons.
            }\hss}}\endlastfoot
2025 Jul 15 12:37:50& $g,r,i$ & 17.4 & 1.68 & \\
12:50:27& $g,r,z$ & 17.4 & 1.67 & Overlap with a bright source\\
13:10:06& $g,r,i$ & 17.4 & 1.66 & \\
13:22:32& $g,r,z$ & 17.4 & 1.67 & \\
13:42:58& $g,r,i$ & 17.4 & 1.69 & Overlap with a faint source\\
13:55:17& $g,r,z$ & 17.4 & 1.71 & \\
14:09:10& $g,r,z$ & 17.4 & 1.74 & Overlap with a faint source \\
14:19:13& $g,r,z$ & 17.4 & 1.77 & \\
14:33:19& $g,r,i$ & 17.4 & 1.83 & \\
14:43:22& $g,r,i$ & 17.3 & 1.87 & Overlap with a faint source \\
14:53:24& $g,r,i$ & 17.3 & 1.92 & \\
15:06:27& $g,r,i$ & 17.3 & 2.01 & Overlap with a bright source\\
\end{longtable}


\section{Results and discussion}\label{sec:res}
\begin{figure}
 \begin{center}
  \includegraphics[width=8cm]{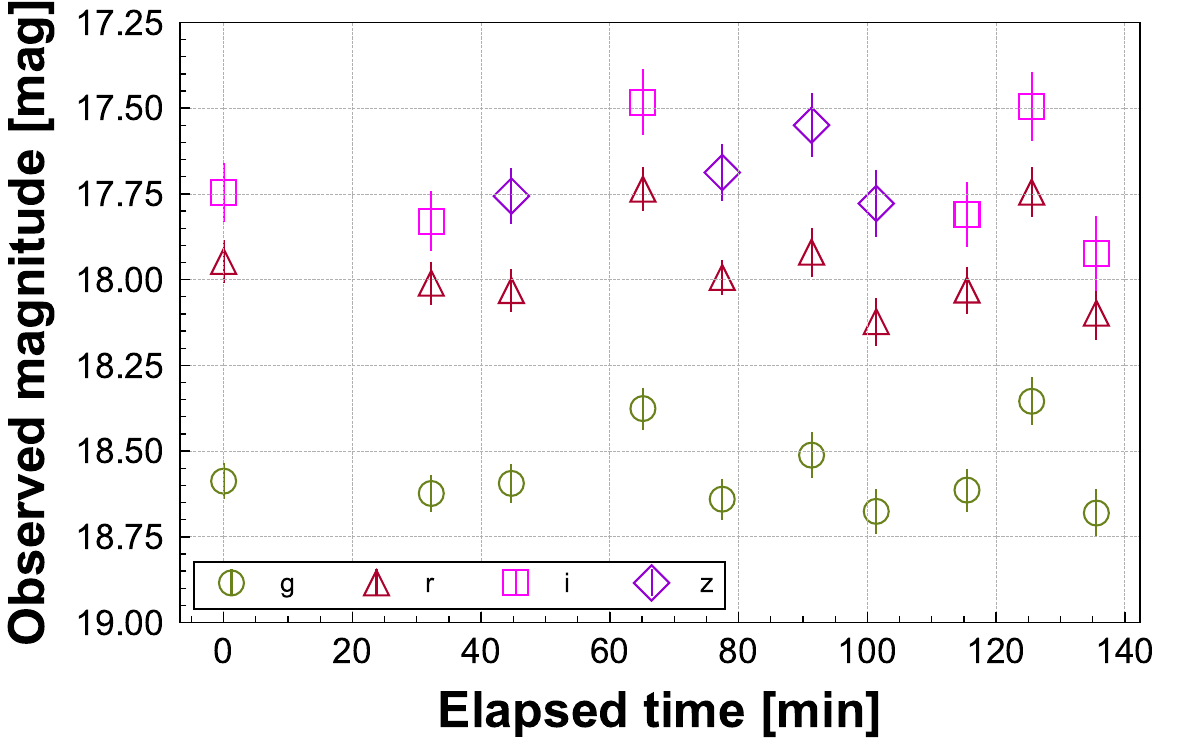} 
 \end{center}
\caption{
    Lightcurves of \I.
    The observed $g$, $r$, $i$, and $z$ bands magnitudes 
    are presented as circles, triangles, squares, and diamonds, respectively.
    Bars indicate the 1$\sigma$ uncertainties.
    {Alt text: Lightcurves of \I in $g$, $r$, $i$, and $z$ bands shown as circles, triangles, squares, and diamonds; error bars show 1$\sigma$ uncertainties.} 
}\label{fig:lc}
\end{figure}

The lightcurves of \I extracted from photometry using a fixed aperture of 6~pix, corresponding to approximately 5000~km from \I, are shown in figure \ref{fig:lc}.
We tested a range of aperture sizes from 5 to 12~pixels and found that the resulting relative brightnesses were nearly identical across this range, except for the largest apertures, where background contamination became non-negligible.
Observations in the $g$ and $r$ bands spanned approximately 2.3~hours, yielding lightcurves that exhibit variations with amplitudes of roughly 0.3~mag.
    It might reflect the rotational variation due to changes in its cross-section. 
    The amplitude of about 0.3~mag is slightly larger than the lightcurve amplitude of about 0.2~mag reported in \citet{Marcos2025_3I},
    while a period of about 2.3~hours is notably short compared to the periodicity of 16.79$\pm$0.23~hours \citep{Marcos2025_3I}. 
    We identified possible contaminations from faint background sources near \I
    by blinking successive images shown in figure \ref{fig:cutout},
    which corresponds to the lightcurve maxima in figure \ref{fig:lc}.
    Therefore, we interpret these brightness variations as likely artifacts rather than rotational variation. 
The relatively flat lightcurves of \I are consistent with \citet{Seligman2025_3I}, 
which found no periodicity is apparent in the intensive time-series observations between July 2 and July 4, 2025. 
\citet{Kareta2025_3I} also reported the absence of rotational variation from photometry spanning over three hours obtained with the IRTF.
    Additional lightcurve observations are crutial for constraining its rotational properties, 
    particularly prior to its perihelion passage in October 2025, when \I is expected to approach the Sun and potentially become more active.

\begin{figure*}
 \begin{center}
  \includegraphics[width=16cm]{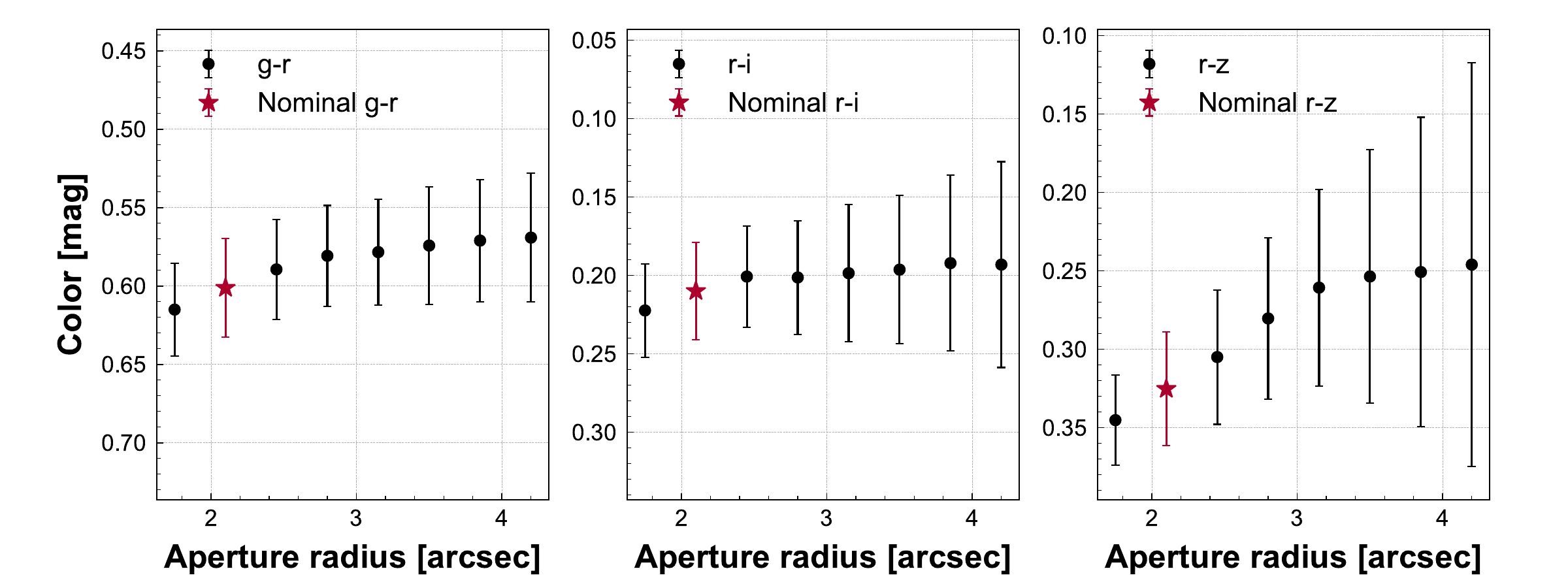} 
 \end{center}
\caption{
    Color indices of \I as a function of aperture radius. 
    The median and standard deviation of the estimated colors are indicated by circles, while the nominal colors are marked with stars.
    Color indices remain nearly constant across aperture sizes from 5 to 12 pixels, indicating robustness of color measurement against aperture size.
    {Alt text: 
    Color indices ($g - r$, $r - i$, $r - z$) of \I vs. aperture radius. Circles indicate median values with standard deviation, stars indicate nominal values; colors remain stable from 5 to 12 pixels.} 
}\label{fig:radcol}
\end{figure*}

As noted above, the lightcurve variations in figure \ref{fig:lc} are likely
instrumental artifacts, while they appear to be synchronized across all bands, suggesting that reliable color measurements can still be obtained.
Colors of \I were computed for each 600-second stacked frame.
We examined a range of aperture sizes from 5 to 12~pixels and found the resulting color indices were identical within the uncertainties as shown in figure \ref{fig:radcol}.
We derived the median values and standard deviations of the color indices to be \gr, \ri, \iz, and \rz using a fixed aperture of 6~pix.
    In table \ref{tab:col}, we summarize the derived colors of \I with those reported in previous studies \citep{Bolin2025_3I, Seligman2025_3I}.
We converted colors from \citet{Bolin2025_3I} in the SDSS system into the Pan-STARRS system using the transformation equations from \cite{Tonry2012}.
We computed the propagated uncertainties of the colors in the Pan-STARRS system with the photometric errors and uncertainties in transformations.

\begin{figure}
 \begin{center}
  \includegraphics[width=8cm]{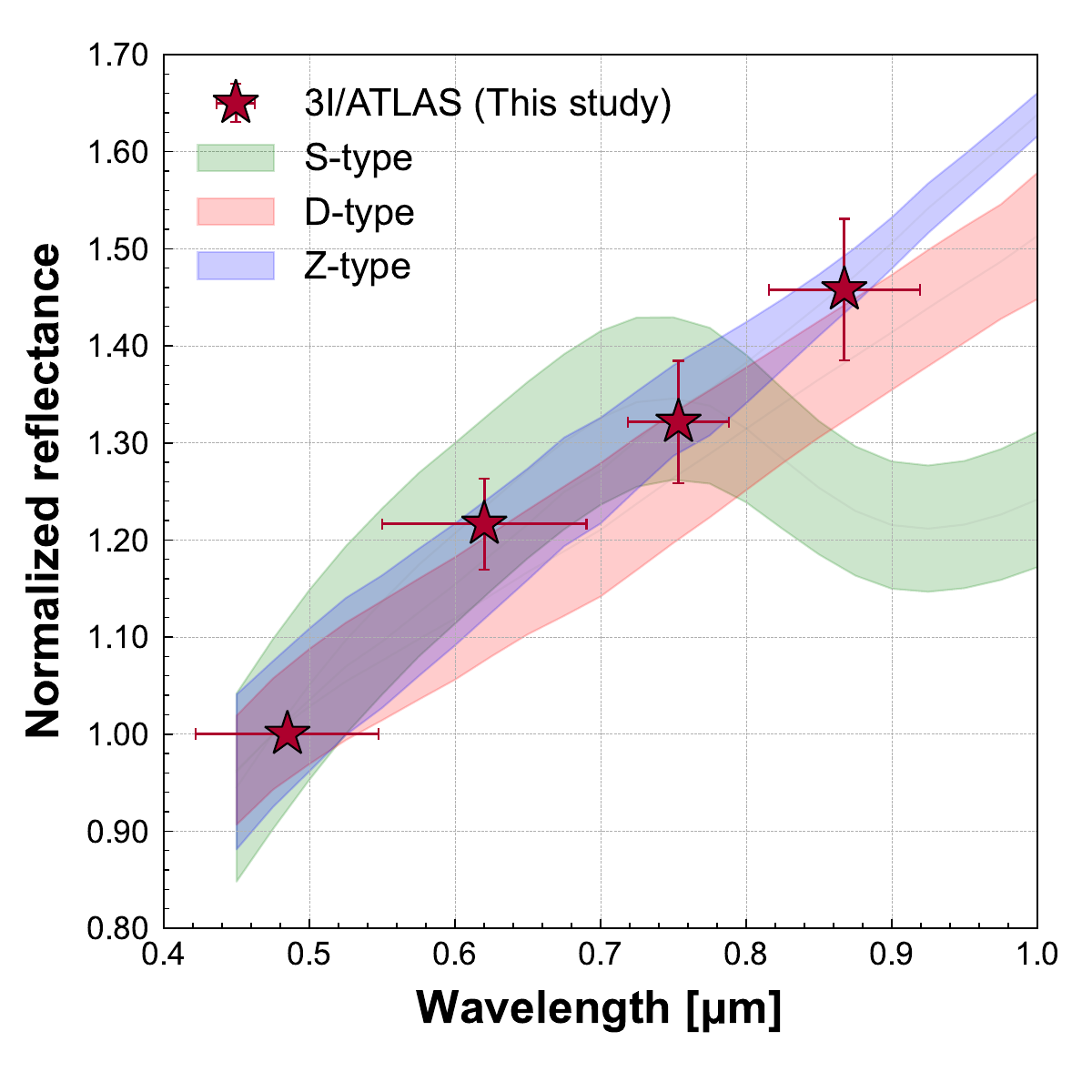} 
 \end{center}
\caption{
Reflectance spectrum of \I.
Also, Mahlke templates of S-, D- and Z-type asteroids are shown \citep{Mahlke2022}.
Shaded areas indicate the standard deviations of the template spectra. 
    {Alt text: Reflectance spectrum of \I compared with S-, D-, and Z-type asteroid templates; shaded areas show 1$\sigma$ variations.} 
}\label{fig:ref}
\end{figure}

In figure \ref{fig:ref}, we show the observed reflectance spectrum of \I based on our visible spectrophotometry.
The reflectances at the central wavelength of the $r$, $i$, and $z$ bands, 
$R_r$, $R_i$, and $R_z$, were calculated as follows \citep[e.g.,][]{DeMeo2013}:
\begin{eqnarray}
    R_r &= 10^{-0.4[(r-g)_{\mathrm{3I}}-(r-g)_\odot]}, \\
    R_i &= 10^{-0.4[(i-g)_{\mathrm{3I}}-(i-g)_\odot]}, \\
    R_z &= 10^{-0.4[(z-g)_{\mathrm{3I}}-(z-g)_\odot]}, 
\end{eqnarray}
where 
$(r-g)_\mathrm{3I}$, $(i-g)_\mathrm{3I}$, and $(z-g)_\mathrm{3I}$ 
are colors of \I,
whereas
$(r-g)_\odot$, $(i-g)_\odot$, and $(z-g)_\odot$ 
are colors of the Sun in the Pan-STARRS system.
We adopted solar magnitudes of $g = 5.03$, $r = 4.64$, $i = 4.52$, and $z = 4.51$ from
\citet{Willmer2018}, and assumed an uncertainty of 0.02 in each.
The reflectance spectra in figure \ref{fig:ref} are normalized at the band center of $g$ band in the Pan-STARRS system, 0.4849~$\mu$m \citep{Willmer2018}.
Horizontal bars in \I's spectrum indicate filter bandwidths.
The calculated reflectance values are 
    $R_r$ of $1.216\pm0.047$, $R_i$ of $1.321\pm0.063$, and $R_z$ of $1.458\pm0.073$.
In figure \ref{fig:ref}, Mahlke templates of S-, D-, and Z-type asteroids \citep{Mahlke2022} are shown for comparison.
The reflectance spectra other than \I are renormalized at the $g$ band center.
We found that the spectrum of \I appears similar to that of D- or Z-type asteroids.

    Our $r-i$, $i-z$ colors indices of \I are 
    consistent with those reported 
    \cite{Seligman2025_3I} and \cite{Bolin2025_3I} within 2$\sigma$.
    The $r-z$ color index of \I is also consistent with that in \cite{Seligman2025_3I} within 1.3$\sigma$, 
    while slightly differs from that in \citet{Bolin2025_3I} at 2.2$\sigma$ level. 
    Our $g-r$ color index of \I differs from 
    \cite{Seligman2025_3I} and \cite{Bolin2025_3I} at 5.8$\sigma$ and 2.1$\sigma$ levels, respectively.
    Our visible reflectance spectrum, which indicates a D-type asteroid surface, is largely in good agreement with the visible spectroscopy \citep{Seligman2025_3I, Opitom2025_3I, Marcos2025_3I, Yang2025_3I}.

    Discrepancies
    in visible colors have previously been reported for 1I/`Oumuamua \citep{Meech2017, Jewitt2017, Bannister2017}.
    \citet{Bannister2017} interpreted that the discrepancy indicates a potential surface color change,
    and this could also be the case for \I.
    Apart from surface color changes, comet spectra may vary due to other mechanisms as well \citep[][and references therein]{Jewitt2015}.
    Possible physical explanations include outbursts
    massive enough to affect the ambient environment \citep[e.g.,][]{Yang2009}
    and chemically heterogeneous dust in the coma \citep{Ivanova2017}.
    \I was at a relatively large heliocentric distance 
    during both our observations and those reported previously \citep{Seligman2025_3I, Bolin2025_3I}, 
    making significant outbursts unlikely.
    No strong activity has been reported to date.
Further characterizations of \I, especially prior to its perihelion passage in October 2025, will be essential to verify these results and investigate physical properties of the interstellar object.

\begin{longtable}{cccccccc}
  \caption{Visible colors of \I in the Pan-STARRS system}\label{tab:col}  
\hline\noalign{\vskip3pt} 
  Reference  & Date & Heliocentric dist.\footnotemark[$*$]  & Aperture\footnotemark[$\dag$]  & $g-r$ & $r-i$ & $i-z$ & $r-z$   \\ [2pt] 
\hline\noalign{\vskip3pt} 
\endfirsthead      
\hline\noalign{\vskip3pt} 
 Reference    & Date (UT) & Heliocentric dist. & Aperture & $g-r$ & $r-i$ & $i-z$ & $r-z$ \\ [2pt]  
\hline\noalign{\vskip3pt} 
\endhead
\hline\noalign{\vskip3pt} 
\endfoot
\hline\noalign{\vskip3pt} 
\multicolumn{2}{@{}l@{}}
{\hbox to0pt{\parbox{160mm}{\footnotesize
\hangindent6pt\noindent
\hbox to6pt{\footnotemark[$*$]\hss}\unskip%
                          Heliocentric distance in au.
    
             {\footnotemark[$\dag$]\hss}\unskip%
            Aperture radius from comet's nucleus in km. 
            
            {\footnotemark[$\ddag$]\hss}\unskip%
            Colors of \I in the SDSS system reported in \cite{Bolin2025_3I} were converted to those in the Pan-STARRS system using the equations in \cite{Tonry2012}. 
}\hss}
} 
\endlastfoot 
  \cite{Seligman2025_3I}                    & 2025 July 2, 4  & 4.47, 4.40     & $\sim$4000     & $0.85\pm0.03$   & $0.25\pm0.03$   & $0.20\pm0.08$    & $0.45\pm0.09$   \\
  \cite{Bolin2025_3I}\footnotemark[$\ddag$]             & 2025 July 6    & 4.33  & $\sim$10000 & $0.717\pm0.044$ & $0.162\pm0.030$ & $-0.018\pm0.070$ & $0.144\pm0.076$            \\
  This study                                         & 2025 July 15   & 4.03  & $\sim$5000  & \grval          &  \rival         & \izval           & \rzval                     \\
\end{longtable}

\section{Conclusion}\label{sec:conc}
We obtained simultaneous visible spectrophotometry of \I using Seimei/TriCCS on July 15, 2025, when \I was at a heliocentric distance of 4.03~au.
The resulting lightcurves exhibit no significant brightness variation over a 2.3-hour time span, consistent with previous reports.
We derived color indices for \I of 
\gr, \ri, \iz, and \rz in the Pan-STARRS system,
suggesting that the surface color of \I closely resembles that of typical D-type asteroids.
Further monitoring observations of \I, 
especially before its perihelion passage in October 2025, will be essential to validate these results and to gain deeper insight into the physical properties of the interstellar object.

\begin{ack}
We thank Dr.~Akito Tajitsu, the Director of Subaru Telescope Okayama Branch Office, 
for the time allocation to observe \I.
We would like to thank Dr. Kenta Taguchi, and Dr. Masaru Kino,
and Dr.~Hideyuki Izumiura for observing assistance using the Seimei telescope.
    We acknowledge the anonymous reviewer for the prompt and constructive comments 
    that improved the quality of the paper.
    The authors thank the TriCCS developer team (which has been
    supported by the JSPS KAKENHI grant Nos. JP18H05223, JP20H00174, and
    JP20H04736, and by NAOJ Joint Development Research)
\end{ack}

\section*{Funding}
This work was supported by JSPS KAKENHI Grant Numbers JP23KJ0640 and 25H00665. 
This work was supported by the French government through the France 2030 
investment plan managed by the National Research Agency (ANR), as part of the
Initiative of Excellence Université Côte d’Azur under reference number ANR- 15-IDEX-01.

\section*{Data availability} 
The data underlying this article are available at the following 
github repository \texttt{https://github.com/jinbeniyama/3I\_Seimei}.


\input{main.bbl}

\bibliographystyle{pasjbib}

\end{document}

%% file: main.bbl
\newcommand{\noopsort}[1]{} \newcommand{\singleletter}[1]{#1}